\documentclass[
    ,final            
  ]
  {aipproc}
\usepackage{amssymb,amsmath}
\layoutstyle{6x9}
\newcommand{\dd}{\mathrm{d}}
\newcommand{\ii}{\mathrm{i}}
\newcommand{\MeV}{\ensuremath{\mathrm{MeV}}}
\newcommand{\GeV}{\ensuremath{\mathrm{GeV}}}
\newcommand{\fm}{\ensuremath{\mathrm{fm}}}

\begin{document}

\title{Heavy Quark Dynamics in the QGP}


\author{V. Greco}{
  address={Department of Physics and Astronomy, University of Catania,}\\ 
\small{\it Via S. Sofia 64, I-95125 Catania (Italy)}}

\author{H. van Hees}{
  address={Institut f\"ur Theoretische Physik, Goethe-Universit\"at Frankfurt,}\\
\small{\it Ruth-Moufang-Str. 1, D-60438 Frankfurt, Germany, }}

\author{R. Rapp}{ 
address={Cyclotron Institute and Physics Department,}\\
\small{\it Texas A{\&}M University, College Station, Texas 77843-3366, U.S.A.}}

\begin{abstract}
  We assess transport properties of heavy quarks in the Quark-Gluon
  Plasma (QGP) that show a strong non-perturbative behavior.  A T-matrix
  approach based on a potential taken from lattice QCD hints at the
  presence of heavy-quark (HQ) resonant scattering with an increasing 
  strength as the temperature, $T$, reaches the critical temperature, 
  $T_c \simeq 170 \; \MeV$ for deconfinement from above.  The 
  implementation of HQ resonance scattering
  along with a hadronization via quark coalescence under the
  conditions of the plasma created in heavy-ion collisions has been
  shown to correctly describe both the nuclear modification factor,
  $R_{AA}$, and the elliptic flow, $v_2$, of single electrons at RHIC
  and have correctly predicted the $R_{AA}$ of D mesons at LHC energy.

\end{abstract}

\maketitle


\section{Introduction}

One of the most interesting questions in high-energy nuclear physics is
about the properties of the hot and dense medium created in
ultra-relativistic heavy-ion collisions. Finite-temperature lattice-QCD
(lQCD) calculations of strong\-ly-in\-ter\-ac\-ting matter predict a
phase transition from hadronic matter to a quark-gluon plasma (QGP) at a
critical temperature, $T_c \simeq 170 \;
\mathrm{MeV}$~\cite{Borsanyi:2010cj}.  The heavy charm and bottom quarks
are particularly valuable probes for the properties of this medium. In
the context of QGP physics they are considered heavy because their
mass, $m_Q$, is large not only with respect to $\Lambda_{\text{QCD}}$
but also to the temperature, $T$, of the plasma. This remains true going
from SPS to LHC energies spanning a $T$ range of $200$-$600 \; \MeV$.
This property makes the study of heavy quark special, because $m_Q \gg
\Lambda_{QCD}$ allows to determine the initial heavy-quark (HQ) spectra 
by means of pQCD and makes available an out-of-equilibrium probe; the
production time $\tau^Q_0 \ll \tau_{\mathrm{QGP}}$ is much smaller than
the QGP lifetime. Therefore heavy quarks pass through the entire evolution 
of the
fireball; the HQ equilibration time $\tau_{\text{eq}}$ is of the
order of the QGP lifetime but smaller than the light-quark one
$\tau^Q_{\mathrm{eq}} \sim \tau_{\mathrm{QGP}} \gg \tau^q_{\text{eq}}$
which means that in principle they carry more information; $m_Q \gg T$
has two important implications: on one hand it imposes effective 
flavor conservation, that in particular holds not only during the QGP
phase but also in the hadronization process; on the other hand it
implies also that the momentum exchange by collisions $|q^2| \ll m_Q^2$
(parametrically dominated by elastic scatterings), and the dynamics can
be treated as a Brownian motion by means of a Fokker-Planck equation
which constitutes a significant simplification of the study of transport
properties. Finally, the three-momentum transfer dominates over energy
transfer $|\vec{q}| \gg q_0 \sim \frac{\vec{q}^2}{m_Q}$ which allows to
use the concept of a potential and therefore to link the HQ physics to
the studies to of the HQ free energy in lQCD~\cite{Petreczky:2004pz}. 
The latter property allows to
employ a finite-temperature $T$-matrix approach to study the problem of
HQ interactions in the medium providing a consistent framework
to evaluate both bound-state and scattering solutions based on a
two-body static potential.

The first results at RHIC further enhanced the potential
interest in HQ dynamics by showing an unexepectedly strong
interaction of heavy quarks with the medium observed through a small 
nuclear modification factor, $R_{AA}(p_T)$, and a large elliptic flow,
$v_2(p_T)$, of the single electrons, $e^\pm$, from semileptonic HQ
decays. Before these experimental results such a behavior was considered
as an unrealistic upper limit, useful only as a 
reference~\cite{Greco:2003vf}.  Instead, the prediction of 
a large $R_{AA}$ and a small $v_2$ obtained from gluon
Bremsstrahlung from heavy quarks has been in striking contrast with the
observations~\cite{Armesto:2005mz,Wicks:2005gt}. Furthemore, it has not
been possible to observe the expected mass hierarchy in the suppression
and its color dependence that would lead to $R_{AA}(B) > R_{AA}(D) >
R_{AA}(h)$ \cite{Armesto:2005mz}. This was in part due to the rather 
indirect experimental access to the HQ dynamics through
the measurements on the single $e^\pm$, which does not allow to
disentagle the individual contributions from $B$ and $D$ mesons. A first
breakthorugh in this direction has been possible thanks to the recent
preliminary results on D mesons presented at QM2011 by the ALICE
Collaboration \cite{dainese:qm2011} which confirm a large suppression of
heavy mesons. In these Proceedings we discuss HQ scattering in
the QGP focusing on the possibility of resonance scattering with light
(anti-)quarks reminiscent of quark confinement. We discuss the
comparison of the model with the data at RHIC and the 2007 prediction
for LHC.

\section{Resonant in-Medium Heavy-Quark Scattering}
Early approaches to the study of HQ scattering in the medium was
carried  our using perturbative QCD (pQCD), first using elastic 
scattering~\cite{Svetitsky:1987gq} and later based on
gluon-brems\-strah\-lung energy loss and/or including elastic HQ
scattering \cite{Armesto:2005mz,Wicks:2005gt}. In such approaches only a
moderate decrease of $R_{AA}$ and a small $v_2$ of the single electrons
from $D$ and $B$ decays have been expected in clear disagreement with
the experimental observations.
Hence non-perturbative approaches are expected to be necessary to
explain the strong HQ interaction with the medium. An early suggestion
postulated a mechanism via the formation of $D$- and $B$-meson resonance 
excitations in the deconfined phase of QCD
matter~\cite{vanHees:2004gq,vanHees:2005wb}. This idea has first been
realized through a non-relativistic effective field theory, modeling
colorless (pseudo-) scalar and (axial-) vector $D$- and $B$-mesons
exploiting both chiral symmetry and HQ (spin) symmetry,
\begin{equation}
\begin{split}
\label{lag-d}
\mathcal{L}_{Dcq} =& \mathcal{L}_D^0 + \mathcal{L}_{c,q}^0 - \ii G_S \left( \bar q \Phi_0^*
\frac{1+{\gamma \cdot v}}{2} c - \bar q \gamma^5 \Phi \frac{1+{\gamma \cdot v}}{2}
 c + h.c. \right) \\
& - G_V \left( \bar q \gamma^{\mu} \Phi_{\mu}^* \frac{1+ \gamma \cdot v}{2} c -
  \bar q \gamma^5 \gamma^{\mu} \Phi_{1\mu} \frac{1+\gamma \cdot v}{2} c + h.c.
\right)
\end{split}
\end{equation}
The fields, ($\Phi^*$) $\Phi$, represent (\emph{anti})-$D$-mesons,
transforming as isospinors under isospin rotations and with the usual
$\mathcal{L}_{c,q}^0$ free terms for quarks and $D$-mesons; see
Ref.~\cite{vanHees:2004gq} for more details.

The interaction terms in Eq.~(\ref{lag-d}) were evaluated to leading
order in $1/m_c$ according to HQ effective theory (HQET). The
main parameter is given by the coupling $G_{S,V}$ varied to allow for
widths of the $D$-meson spectral functions of $300$-$500 \; \MeV$, to
approximately cover the range suggested by effective quark models.  It
is important to note that we assume the $D$-meson resonances, $m_D=2.0$
GeV to be located above the $c-\bar{q}$ mass threshold, $m_c+m_q=1.5$
GeV, which renders them accessible in $c-\bar{q}$ scattering
processes. The situation is quite different for (bound) meson states
(i.e., below the anti-/quark threshold), where the resonant part of the
scattering amplitude cannot be probed through $c+\bar{q} \rightarrow
c+\bar{q}$ interactions (even for resonance masses close to threshold,
thermal energies of anti-/quarks imply that the average collision energy
is significantly above the resonance peak).

We find that the presence of these resonances at moderate QGP
temperatures substantially accelerates the kinetic equilibration of
$c$-quarks as compared to using perturbative interactions. We have
concentrated on the charm-quark case, but completely analogous
expressions apply to the bottom sector. These approaches have been used
in 2006 to make predictions for the case of Au+Au at RHIC energies, and
the details are presented in Ref.~\cite{vanHees:2005wb}. The key
ingredients are that the transport cross section is appreciably larger 
than the pQCD ones and a hadronization mechanism that includes quark
coalescence~\cite{Greco:2003mm,Greco:2007nu,Fries:2008hs},
leading to an enhancement of both the quark $R_{AA}$ and $v_2$ toward a
better agreement with the data \cite{vanHees:2005wb,vanHees:2007me}.
Other approaches have consistently pointed out the necessity to go
beyond a simple pQCD scheme \cite{Gossiaux:2008jv,Alberico:2011zy} to
account for the observed $R_{AA}$ even if these studies included 
coalescence especially in the $p_T^{e}\leq 3$-$4 \; \GeV$
region, which translates to $p_T^D \leq 7$-$8 \; \GeV$ at the meson
level.  In particular our approach appeared to be the only one able to
describe both the small $R_{AA}$ and large $v_2(p_T)$ simultaneuosly. 
Furthermore a
prediction for Pb+Pb collisions at LHC energy has been presented in the
last call for prediction proceedings \cite{Abreu:2007kv} and will be
discussed in the following.

\subsection{T-matrix Approach to Heavy-Quark Scattering}

The success of the first applications of the idea of resonant HQ 
scattering in the medium has lead to a further and more realistic
assessment of the existence of resonance scattering. The idea of the
existence of resonance scattering is indeed supported by lQCD on the
quark correlators for both heavy and light quarks \cite{AH-prl} as well
as by Non-Relativsitic QCD solved on lattice for heavy quarks \cite{Aarts:2010ek}. They
both show the existence of a peak in the spectral function even at
temperatures substantially higher than $T_c$ suggesting the presence of
a physical mechanism beyond a simple free scattering.

As mentioned in the introduction, the large HQ mass allows the
use of an interaction potential between quarks. This has the advantage
that $T$-matrix scattering theory becomes applicable, which does not
rely on a perturbative expansion but is able to account also for
moderate or even strong coupling, where resummations of large diagrams
are necessary and realized via the standard ladder sum.  An extra
benefit is that one can in principle extract the quark potential from
finite-temperature lattice QCD (lQCD), or at least be constrained by
lQCD ``data'' which gives a parameter-free input.

A static heavy-quark light-quark potential, $V(r)$, has been used in the
vacuum to successfully describe $D$-meson spectra and
decays~\cite{Godfrey:1985xj,Avila:1994vi}. We assume that the effective
in-medium potential can be extracted from finite-temperature lQCD
calculations of the color-singlet free energy
$F_1(r,T)$~\cite{Kaczmarek:2003dp,Kaczmarek:2005gi} for a static
$\overline{Q} Q$ pair as the internal potential energy by the usual
thermodynamic
relation~\cite{Mannarelli:2005pz,Shuryak:2004tx,Wong:2004zr},
\begin{equation}
\label{F-to-U}
U_1(r,T)=F_1(r,T)-T \frac{\partial F_1(r,T)}{\partial T}.
\end{equation}
For the application as a scattering kernel in a $T$-matrix equation, the
potential has to vanish for $r \rightarrow \infty$. Thus we choose the
accordingly subtracted internal potential energy,
\begin{equation}
\label{U-to-V}
V_1(r,T)=U_1(r,T)-U_1(r \rightarrow \infty,T).
\end{equation}
In lQCD simulations one finds that $U_1(r \rightarrow \infty,T)$ is a
decreasing function with temperature which could be associated as a
contribution to the in-medium HQ mass, $m_Q(T)=m_0+U_1(r \rightarrow
\infty,T)/2$, where $m_0$ denotes the bare mass. However, 
close to $T_c$ the asymptotic value, $U_1(r \rightarrow
\infty,T)$, develops a pronounced peak structure which is currently
not fully understood (possibly related to multiple coupled channel
effects). For simplicity,  in the current
calculation, we assume constant effective HQ masses, $m_c=1.5 \;
\mathrm{GeV}$ and $m_b=4.5 \; \mathrm{GeV}$.

We also consider the complete set of color channels for the $Q \bar{q}$
(singlet and octet) and $Q q$ (anti-triplet and sextet) systems, using
Casimir scaling as in leading-order pQCD, $ V_{8}=-V_1/8, \quad
V_{\bar{3}}= V_1/2., \quad V_6=- V_1/4$, which is also justified by lQCD
calculations of the finite-$T$ HQ free energy~\cite{Doring:2007uh}.

The uncertainties due to the extraction of the potential have a moderate
final impact on the agreement with the experimental data, see also
Fig.\ref{fig:raa-v2-el} (left), while the difference in the transport
coefficients going from the internal energy, $U$, to the free energy,
$F$, is quite substantial.  However, the successful application to
compute quarkonium correlators and HQ susceptibilities lends
a-posteriori support (albeit not validation) for the choice of $U$,
which is not as convincing for $F$ (cf.~Ref.~\cite{Riek:2010fk}).

\begin{figure}
  \centerline{\includegraphics[height=.23\textheight]{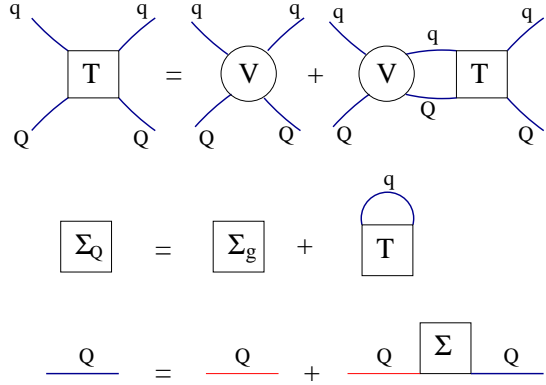}}
  \caption{Diagrammatic representation of the Brueckner many-body scheme
    for the coupled system of the $T$-matrix based on lQCD static internal
    potential energy as the interaction kernel and the HQ self-energy.}
\label{fig_BSSD}
\end{figure}

The starting point of HQ Brueckner theory is a system of coupled
Bethe-Salpeter (BS) and Schwinger-Dyson (SD) equations characterizing
the HQ interaction and propagator in the QGP,
\begin{equation}
\label{SD}
  M = K + \int K
  G M, \quad
  \Sigma^Q = \Sigma^Q_g +\int\! T S^q, \quad
  S^Q = S^Q_{0} + S^Q_{0} \Sigma^Q S^Q \, ,  
\end{equation}
where $M$ denotes the scattering amplitude between a heavy ($Q$) and a
light ($q$) quark or antiquark, $K$ the two-body interaction kernel, $G$
the two-particle ($qQ$) propagator, $S^{Q,q}$ ($S_0^{Q,q}$) the (free)
single-particle propagator, and $\Sigma^Q$ the HQ selfenergy receiving
contributions from thermal gluons ($\Sigma_g$) and light quarks (where
the latter are computed self-consistently from the heavy-light
scattering amplitude). Since we focus on a QGP at zero chemical
potential ($\mu_q$=0), all quantities are quark-antiquark symmetric. The
predominantly space-like momentum transfer in on-shell scattering of
heavy quarks, $q^2=q_0^2-\vec q^{\,2} \simeq -\vec q^{\,2}$, justifies a
static (potential) approximation to its interaction, $K\approx V$.  This
allows to reduce the four-dimensional (4D) BS equation into a 3D
Lippmann-Schwinger equation for the $T$-matrix, which greatly simplifies
its solution. Using azimuthal symmetry and a partial-wave expansion
leads to a one-dimensional integral equation for the amplitudes, $T_l$,
of given angular momentum, $l$,
\begin{equation}
\label{Tmat}
T_{l}^a(E;p^\prime, p) = V_l^a(p^\prime, p) + 
                     \frac{2}{\pi}\int \dd k \ k^2 \ V_l^a(p^\prime,k) 
                    \ G_{Qq}(E;k) \ T_l^a(E;k,p) \ , 
\end{equation}
where we also indicated the four possible color channels, $a$, for
$Q\bar q$ (singlet and octet, $a=1$ and 8) and $Qq$ (antitriplet and
sextet, $a=\bar{3}$ and 6) states.

The in-medium HQ quasiparticle-dispersion relation is given by
\begin{equation}
\label{disp}
\omega^Q_k = \sqrt{k^2+m_Q^2(T)} + {\rm Re}\,\Sigma^Q_q(\omega_k^Q,k), 
\end{equation}
where it has been assumed that Eq.~(\ref{SD}) (middle) can be decomposed
into two distinct contributions: a ``gluon-induced'' one, $\Sigma_g^Q$,
generating a temperature-dependent mass, $m_Q(T)$, and a selfenergy,
$\Sigma^Q_q$, due to scattering of heavy quarks off thermal light
quarks. The former is associated with the long-distance limit of the
potential, while the latter is explicitly evaluated from the above
heavy-light $T$-matrix.  To close the system of equations~(\ref{SD}) in
the quark sector, the analogous system for the light sector is required,
which has been solved selfconsistently for $\Sigma^q$ and $T_{qq,\bar
  qq}$ in Ref.~\cite{Mannarelli:2005pz}. Guided by the results obtained
there we employ a light-quark propagator (figuring into the second term
in Eq.~(\ref{SD}) with a constant thermal light-quark mass,
$m_q=0.25\; \GeV$, and width, $\Gamma_q=200 \; \MeV =-2 \Sigma^q$.

Clearly, the potential approximation is less reliable in the light-quark
sector, but it turns out that their in-medium selfenergies (real and
imaginary parts), which are needed for the HQ selfenergy,
Eq.~(\ref{SD}), have a small effect on both the scattering amplitude,
$T_{Qq}$, and the HQ transport coeffcients.
 
The above system of equations (for $T$ and $V$) is pictorially
represented in Fig.~\ref{fig_BSSD} with the upper, middle and lower
panel corresponding to Eqs.~(\ref{SD}), respectively.

We restrict ourselves to $S$ ($l=0$) and $P$ ($l=1$) waves.  As can be
seen from Fig.~\ref{fig.frict-coeff} (left), in the dominating
attractive color-singlet $Q\bar{q}$ and color-antitriplet $Q q$
channels, close to the critical temperature, $T_c$, resonance states
close to threshold, $E_{\text{thr}}=m_Q+m_q$ are formed, similar as
conjectured in the effective resonance model described in the previous
section ~\cite{vanHees:2004gq,vanHees:2005wb}. However, in this full
in-medium scheme the resonances melt at lower temperatures $T \gtrsim
1.7 \, T_c$ and $T \gtrsim 1.4 \, T_c$, respectively.

\section{Heavy-Quark observables in Heavy-ion collisions}

A direct comparison between the microscopic description of the HQ
dynamics in the QGP and the experimental observables necessitates a
dynamical implementation of the HQ scattering in the medium plus a model
for the hadronization and finally the implementation of the semileptonic
decay.

\begin{figure}
\includegraphics[width=0.32\textheight]{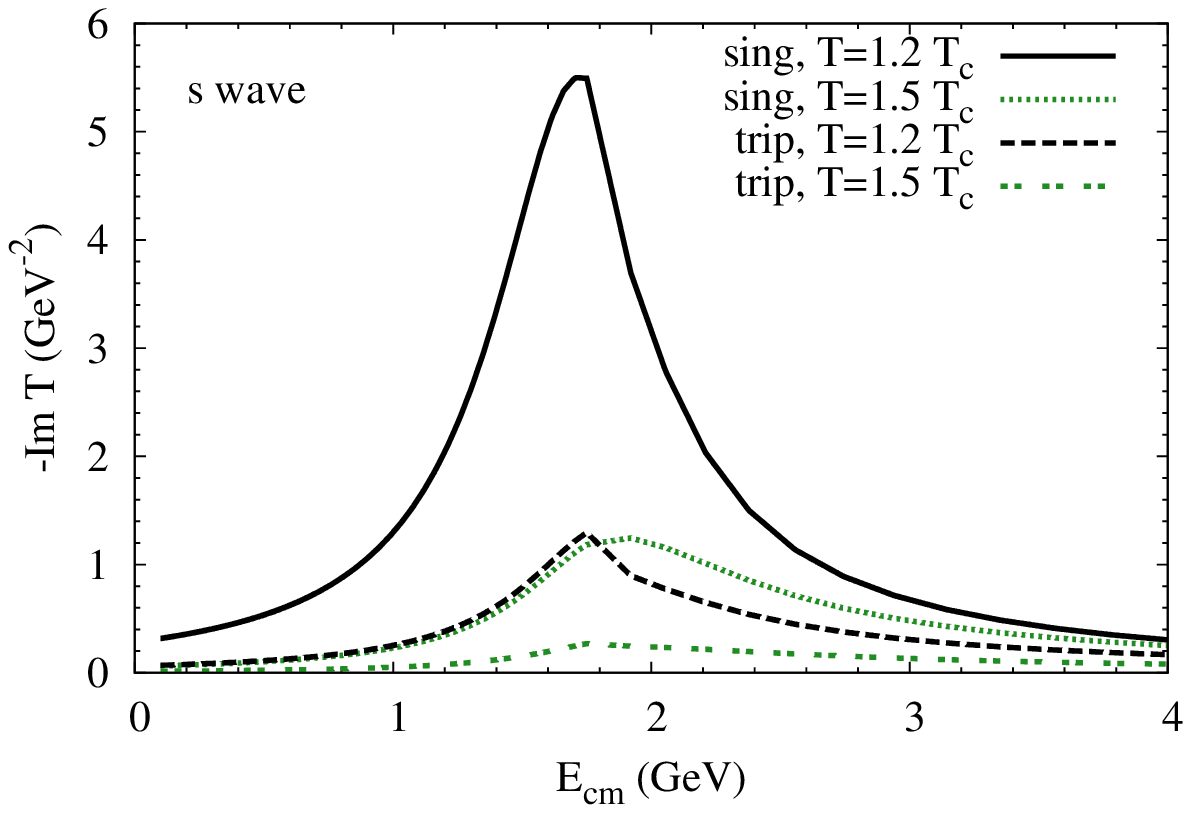}
\includegraphics[width=0.3\textheight]{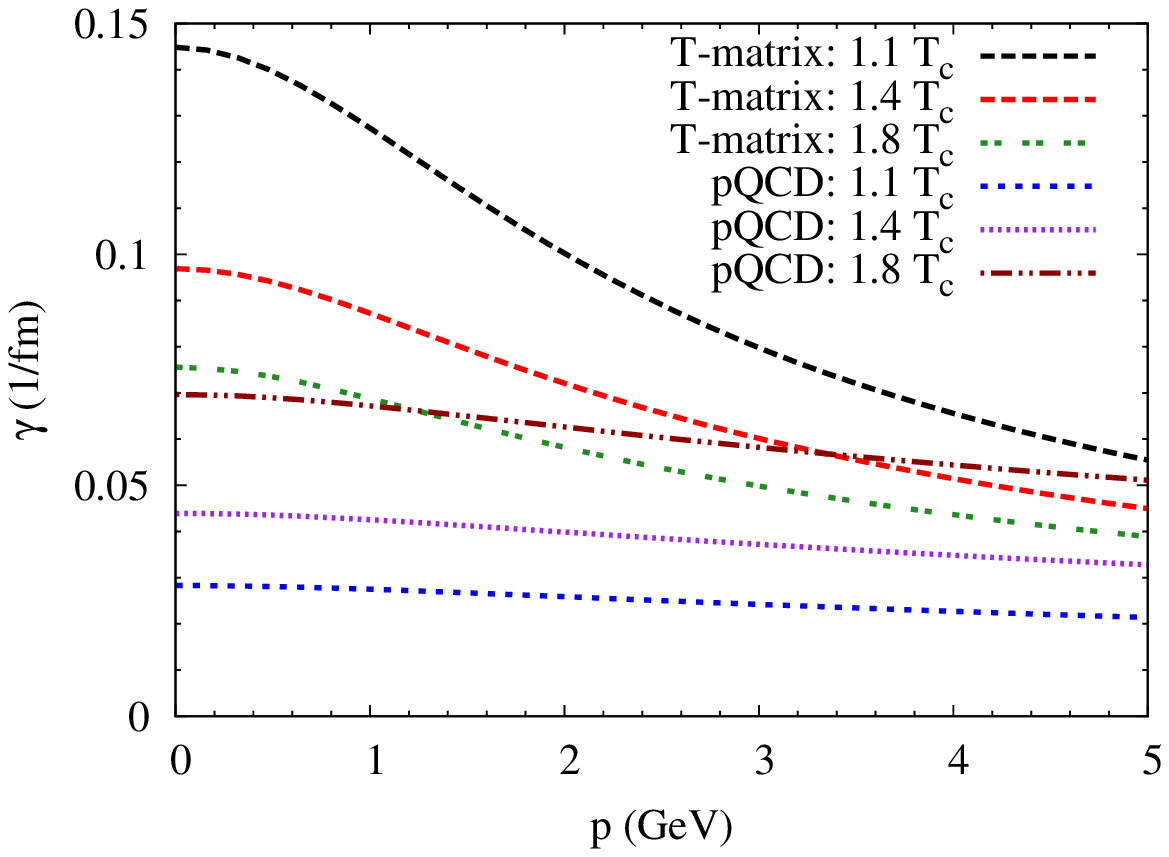}
\caption{Left: Imaginary part of the $S$-wave in-medium $T$ matrix for
  $c\bar{q}$ and $cq$ scattering in the color-singlet and -antitriplet
  channels based on the parameterization of the lQCD potential energy by
  [Wo]. Right: The drag coefficient, $\gamma$, as a function of HQ
  momentum, calculated via (\ref{SD}) with scattering-matrix elements
  from the non-perturbative $T$-matrix calculation (using the
  parameterization of the lQCD internal potential energies by [Wo])
  compared to a LO perturbative calculation based on matrix elements.}
\label{fig.frict-coeff}
\end{figure}

The HQ motion in
the hot and dense QGP, consisting of light quarks and gluons, can be
described by a Langevin simulation of the Fokker-Planck equation,
\begin{equation}
\label{FP}
\frac{\partial f_Q}{\partial t}=\frac{\partial}{\partial p_i} (p_i \gamma
f_Q) + \frac{\partial^2}{p_i p_j} (B_{ij} f_Q).
\end{equation}
The drag or friction coefficient, $\gamma$, and diffusion coefficients,
$B_{ij}$, are calculated from the invariant scat\-te\-ring-ma\-trix
elements~\cite{Svetitsky:1987gq}. Taking into account elastic scattering
of the heavy quark with a light quark or antiquark given in terms of the
above calculated $T$-matrix one can calculate the drag with a standard 
procedures the HQ drag and diffusion coefficients, shown in 
Fig.~\ref{fig.frict-coeff}, see Ref.~\cite{Rapp:2008tf} for more details.



The nonperturbative HQ light-quark scat\-te\-ring-ma\-trix ele\-ments
are supplemented by the corresponding perturbative elastic HQ
gluon-scattering ones. The $t$-channel singularity is regulated by a
gluon-Debye screening mass of $m_g=g T$ with a strong coupling constant,
$g=\sqrt{4 \pi \alpha_s}$, using $\alpha_s=0.4$.

As shown in Fig.~\ref{fig.frict-coeff}, close to $T_c$ the equilibration
times,  $\tau_{\text{eq}}=1/\gamma \simeq 7 \; \text{fm}/c$, for charm
quarks are a factor of $\sim 4$ smaller than for a
corresponding pQCD calculation, reminiscent to the results based on the
model assuming $D$-meson like resonance states above
$T_c$~\cite{vanHees:2004gq,vanHees:2005wb}. In contrast to this and
other calculations of the HQ transport coefficients, here the drag
coefficients \emph{decrease} with increasing temperature because of the
``melting'' of the dynamically generated resonances at increasing
temperatures due to the diminishing interaction strength from the lQCD
potentials.

To solve the Fokker-Planck equation~(\ref{FP}) under conditions of the
sQGP medium produced in heavy-ion collisions, we use an isentropically
expanding thermal fireball model, assuming an ideal-gas equation of
state of $N_f=2.5$ effective massless light-quark flavors and gluons.
The initial spatial distribution of HQ production is determined with a 
Glauber model.  The $c$-quark spectra are taken from a modified
PYTHIA calculation to fit $D$ and $D^*$ spectra in d-Au
collisions~\cite{Adams:2004fc}, assuming $\delta$-function
fragmentation. The $b$-quark $p_T$ spectrum is taken from PYTHIA
assuming a cross-section ratio of $\sigma_{b\bar{b}}/\sigma_{c\bar{c}}
\simeq 5 \cdot 10^{-3}$ and a crossing of the $c$- and $b$-decay
electron spectra at $p_t \simeq 5\;\text{GeV}$, consistent with FONLL 
pQCD calculations \cite{Cacciari:2005rk}.

The last step toward a comparison of the above described mo\-del for HQ
diffusion in the QGP with the $e^\pm$ data from RHIC is
the hadronization of the HQ spectra to $D$- and $B$-mesons and their
subsequent semileptonic decay. Here we use the
quark-coalescence model described
in~\cite{Greco:2003mm,Greco:2007nu}. In recent years, the coalescence of
quarks in the hot and dense medium created in heavy-ion collisions has
been shown to provide a successful ha\-dro\-ni\-za\-tion mechanism to
explain phenomena such as the scaling of hadronic elliptic-flow
parameters, $v_2$, with the number of constituent quarks, $v_{2,h}(p_t)
= n_h v_{2,q}(p_t/n_h)$, where $n_h=2(3)$ for mesons (hadrons) denotes
the number of constituent quarks contained in the hadron, $h$, and the
large $p/\pi$ ratio in Au-Au compared to $p$-$p$
collisions~\cite{Greco:2003mm,Fries:2003kq,Greco:2007nu}. Quark
coalescence is most efficient in the low-$p_T$ regime where most $c$ and
$b$ quarks combine into $D$ and $B$ mesons, respectively. To conserve
the total HQ number, we assume that the remaining heavy quarks hadronize
via ($\delta$-func\-tion) fragmentation.

\begin{figure}
  \includegraphics[height=.29\textheight]{RAA-v2-elec-tmat.eps} \hfill
  \includegraphics[height=.29\textheight]{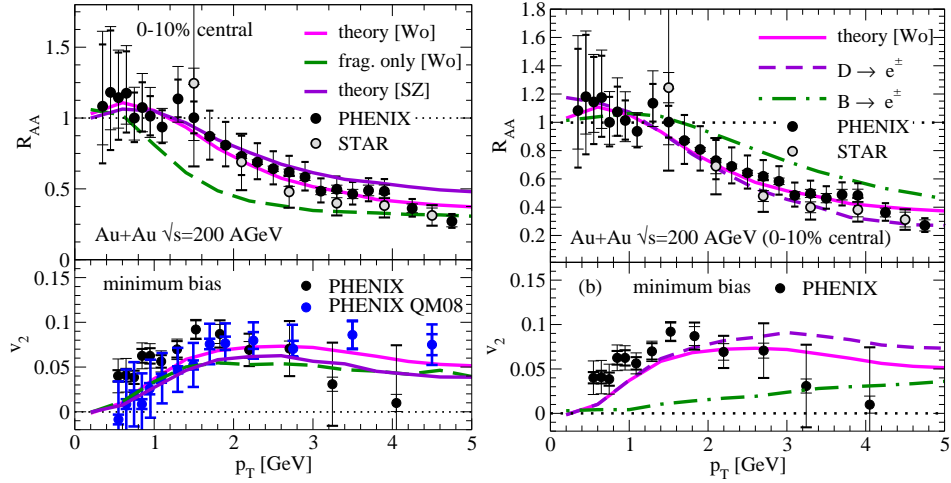}
  \caption{Left: Results for the nuclear modification factor (upper
    panel) and elliptic flow (lower panel) of single electrons
    with/without (solid/dashed lines) quark coalescence in Au-Au
    collisions compared to RHIC
    data~\cite{Abelev:2006db,Adare:2006nq}. Right: As in the left panel
    but only for the case of [Wo] potential with coalescence included it
    is shown the contribution of single $e\pm$ from D and B decay.}
  \label{fig:raa-v2-el}
\end{figure}

As shown in Fig.~\ref{fig:raa-v2-el} the Langevin simulation of the HQ
diffusion, followed by the combined quark-coa\-le\-scen\-ce
fragmentation description of hadronization to $D$ and $B$ mesons and
their subsequent semileptonic decay, successfully accounts
for both the $R_{AA}$ and $v_2$ of $e^\pm$ in
$200 \; A\text{GeV}$ Au-Au collisions~\cite{Adare:2006nq,Abelev:2006db}
at RHIC. The uncertainty due to two different parameterizations of
the lQCD potentials by [Wo] \cite{Wong:2004zr} and [SZ] \cite{Shuryak:2004tx}
is not so large.

Comparing the solid and dashed lines one sees that the effect from the
``momentum kick'' of the light quarks in coalescence, an
enhancement of both $R_{AA}$ and $v_2$, is important for the 
agreement of both observables with the data. As can be seen from the
lower panel in Fig.~\ref{fig:raa-v2-el} (right), the
effects of the $B$-meson decay contribution to the
$e^{\pm}$ spectra become visible for $p_T \ge 2.5$-$3
\; \text{GeV}$. A closer inspection of the time evolution of the $p_t$
spectra shows that the suppression of high-$p_T$ heavy quarks occurs
mostly in the beginning of the time evolution, while the $v_2$ is built
up later at temperatures close to $T_c$ which is to be expected since
the $v_2$ of the bulk medium is fully developed at later stages only
\cite{Rapp:2008zq,Scardina:2010zz}. 
This effect is more pronounced due to resonance formation, 
because the transport coefficients become larger close to $T_c$, or 
at least decrease much slower than in the pQCD case~\cite{Riek:2010fk}.

\subsection{Prediction at LHC}

In 2007 we made predictions for the the $D$ and $B$ spectra at
the LHC employing the effective resonant model described above.
For $D$ mesons we know that the model is quite reliable in the sense
that it generates predictions similar to the $T$-matrix approach.  Since
the initial temperatures at the LHC are expected to exceed the resonance
dissociation temperatures, the prediction implemented a
switching off of the resonances at $T_{\text{diss}}=2\,T_c=360 \; \MeV$
by a factor $(1+\exp[(T - T_{\text{diss}})/\Delta])^{-1}$ with
($\Delta=50 \; \MeV$) in the transport coefficients.

The temperature evolution in the fireball assumed a total entropy fixed by 
the number of charged hadrons which we have extrapolated to $\dd N_{ch}/\dd
y \simeq 1400$ for central $\sqrt{s_{NN}}=5.5 \; \text{TeV}$ Pb-Pb
collisions leading to an initial temperature, $T_0 \simeq 520 \; \MeV$.
This multiplicity truns out to be very close to the
one measured at $\sqrt{s_{NN}}=2.75$ TeV, hence now we know that these
have to be considered more properly as predictions for this energy.

Initial HQ $p_T$ spectra are computed using PYTHIA with parameters as
used by the ALICE Collaboration \cite{Alessandro:2006yt}. Hadronization
is treated as previously discussed for the RHIC case. The shadowing has
not been included but as shown by the orange solid line this should not
affect the prediction at $p_T > 3.5 \;\GeV$.

\begin{figure}
  \includegraphics[height=.28\textheight]{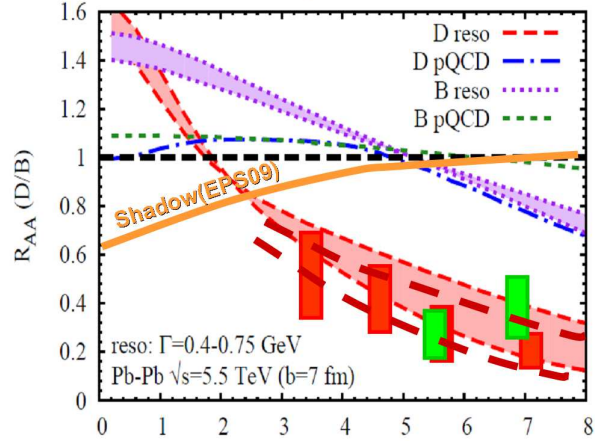}
  \caption{The nuclear modification for D meson at LHC for $Pb+Pb$ at
    $\sqrt{s_{NN}}=2.75$ TeV for more detail see the text.}
\label{fig:D-LHC}
\end{figure}

In Fig.~\ref{fig:D-LHC} the predictions for the $D$-meson $R_{AA}$ are
shown by the shaded red area corresponding to the uncertainties in the
resonance model. The data from the ALICE Collaboration
\cite{dainese:qm2011} are shown by rectangles, red for $D^+$ and green
for $D^0$, and are for the $0-20\%$ centrality and therefore
corresponding to an impact parameter, $b=4.7 \; \fm$, hence more central
with respect to the 2007 calculations. For this reason in the figure the
red dashed line is drawn to indicate the extrapolation from the $b=7 \;
\fm$ calculation to the more central with $b=4.7 \; \fm$ which results
in a $15\%$ correction. It is clear that the prediction from simple pQCD
elastic scattering, shown by the blue dash-dotted line, cannot account
at all for the observed suppression of the $D$ spectra.

Data with a better statistics and especially a measurement of the elliptic 
flow will greatly improve the discrimination power of the model. From 
the theorethical side
it is desireable to have predictions using the $T$-matrix approach based
on the lQCD-based  potential, especially for the $B$ mesons. These
are differently affected by the medium and should show a resonant
scattering persistent up to higher temperatures compared to the $D$
mesons resulting in a smaller $R_{AA}$ compared to the simplified
resonance model shown by the violet shaded area shown in
Fig. \ref{fig:D-LHC}.

\section{Conclusions}

The in-medium interaction of heavy quarks have been intriguing  
from the first preliminary results at RHIC showing a much 
larger suppression than expected, together with a large
elliptic flow, as inferred from semileptonic HQ decay electrons.
An in-medium T-matrix approach (utilizing
potentials estimated from lattice QCD) has been applied to evaluate HQ
interactions in the QGP showing the existence of in medium prehadronic
and diquark resonance states, which increase in strength when
approaching $T_c$.
When implemented into Langevin simulations at RHIC, reasonable agreement
with both the suppression and elliptic flow of $e^\pm$ spectra from HQ
decays emerges.
 
Several theoretical problems remain open like a proper definition of the
potential, corrections to the $T$-matrix approach including radiative
ones, in-medium mass and width effects as well as a self-consistent
treatment of the hadronization via coalescence.  On the other hand this
field is entering a new stage thanks to the possibility to disentangle
the $B$ and $D$ mesons which in itself can provide the key to
discriminate different models
\cite{vanHees:2007me,Gossiaux:2008jv,Alberico:2011zy,Uphoff:2011ad,He:2011qa} .

\begin{theacknowledgments}

  VG is supported by the MIUR under
  the Firb Research Grant RBFR0814TT and by the ERC-StG2010 under the
  QGPDyn Grant n.259684. RR is supported by the US National Science Foundation under
  grant no.~PHY-0969394 and by the A.-v.-Humboldt foundation.

\end{theacknowledgments}


\bibliographystyle{aipproc}   

\end{document}